\documentclass[10pt,journal,twocolumn,twoside]{IEEEtran} 
\usepackage{graphicx}
\usepackage{epstopdf}
\usepackage{float}
\usepackage{algorithmic}
\usepackage{array}
\usepackage{amsmath}
\usepackage{amssymb}
\usepackage{mdwmath}
\usepackage{booktabs}
\usepackage{eqparbox}
\usepackage{stfloats}
\usepackage{tabularx}
\usepackage{hyperref}
\usepackage{cleveref}
\hypersetup{hypertex=true,
            colorlinks=true,
            linkcolor=blue,
            anchorcolor=blue,
            citecolor=blue}
\usepackage{cases} 
\usepackage{subfigure}
\usepackage{upgreek}
\usepackage{multirow}
\usepackage{makecell}
\usepackage{amsmath}
\usepackage[boxed,ruled,commentsnumbered]{algorithm2e}
\usepackage{url}
\usepackage[table,xcdraw]{xcolor}
\usepackage{colortbl}
\usepackage{cite}
\usepackage{xcolor}
\ifCLASSOPTIONcompsoc
\usepackage[caption=false,font=normalsize,labelfont=sf,textfont=sf]{subfig}
\else
\fi
\allowdisplaybreaks[4]

\makeatletter

\renewcommand*{\@opargbegintheorem}[3]{\trivlist
      \item[\hskip \labelsep{\bfseries #1\ #2}] \textbf{(#3):}\ }
\makeatother

\begin{document}
\title
{Structured Pose-Conditioned Flow Matching for Generative 5G CSI Augmentation}

\author{Haojin Li, Anbang Zhang, 
Wai Ho Mow,~\IEEEmembership{Senior Member, IEEE}, Chenyuan Feng,~\IEEEmembership{Senior Member,~IEEE},\\ Chen Sun,~\IEEEmembership{Senior Member, IEEE}, and Haijun Zhang,~\IEEEmembership{Fellow, IEEE} 
\thanks{
(\emph{*Corresponding author: Anbang Zhang})

Haojin Li and Haijun Zhang are with University of Science and Technology Beijing, China (email: Haojin.li@sony.com, haijunzhang@ieee.org). 

Haojin Li and Chen Sun are with Sony China Research Laboratory, China (email:  chen.sun@sony.com).

Anbang Zhang and Wai Ho Mow are with the Department of ECE, The Hong Kong University of Science and Technology, Hong Kong (e-mail: azhangay@connect.ust.hk, eewhmow@ust.hk).

Chenyuan Feng is with University of Exeter. (email: c.feng@exeter.ac.uk).
}}
\maketitle


\begin{abstract} 

With the growing demand for privacy-preserving and occlusion-resilient human pose recognition (HPR), 5G channel state information (CSI) offers a promising contactless sensing modality by integrating communication and sensing capabilities. However, collecting large-scale synchronized CSI-pose pairs remains costly in practical 5G systems. To address this limitation, we propose StructFlow-HPR, a structured pose-conditioned flow matching framework for generative CSI augmentation. StructFlow-HPR learns a continuous latent transport process from Gaussian noise to real CSI representations under pose guidance, while preserving the receiver-frequency topology of CSI through a reconstruction-preserving autoencoder. A pose-conditioned Transformer is further designed to model the latent velocity field and generate pose-aligned CSI samples via ordinary differential equation sampling. Experiments on real-world 5G sensing data show that StructFlow-HPR can produce realistic CSI-pose pairs and improve downstream HPR performance under limited-data conditions.

\end{abstract}


\section{Introduction}

\IEEEPARstart{W}{ith} the evolution of ubiquitous computing and spatial intelligence, high-precision human pose recognition (HPR) has emerged as a key capability for connecting the physical world with digital twin spaces \cite{10944626}. Traditional mainstream sensing paradigms rely heavily on optical vision sensors (e.g., RGB and depth cameras) \cite{vitposepp2024tpami}. Admittedly, vision-based models possess intuitive advantages in capturing fine-grained skeletal keypoints across continuous video sequences \cite{dwpose2023iccvw}. 
However, large-scale deployment in indoor scenarios is limited by line-of-sight (LoS) conditions and is susceptible to illumination changes and physical occlusions \cite{humanart2023cvpr}.

To overcome the limitations of optical sensing, contactless radio frequency (RF) sensing technologies have emerged as a viable alternative for through-wall and occlusion-resistant sensing \cite{personinwifi3d2024cvpr}. 
Specifically, 5G communication networks, owing to the native integrated sensing and communication (ISAC) \cite{11145172} architecture, can exploit channel state information (CSI) to accurately capture human motion-induced variations.

Unlike bandwidth-limited WiFi protocols \cite{capshar2024jiot}, which often require specialized attention mechanisms \cite{10152057} and distinct from radar systems requiring dedicated hardware arrays,
5G CSI encapsulate exceptionally rich features of high-frequency spatial multiplexing and multipath fading \cite{isac_standardization2024mcomstd}. By resolving the microscopic phase shifts and Doppler perturbations during the reflection and scattering of 5G signals off the human body, the system can provide fine-grained spatial resolution and ultra-low latency, thereby enabling high-dimensional mapping of continuous human kinematic states.

However, precisely mapping high-dimensional and complex RF physical signals into the 3D human kinematic space heavily relies on the powerful nonlinear fitting capabilities of deep neural networks. In practical deployment, this paradigm encounters severe challenges of data scarcity and severe overfitting \cite{diffusionts2024iclr}. 
In real-world indoor 5G communication environments, the time and labor costs associated with synchronously acquiring and annotating large-scale, high-precision data are prohibitively expensive. 
Thus, generative data augmentation emerges as a highly promising breakthrough pathway in modern wireless networks \cite{dit2023iccv}.
Rather than fitting models directly to limited raw datasets, it is more effective to synthesize massive volumes of virtual training data to overcome indoor localization and recognition bottlenecks \cite{flowmatching2023iclr}. By learning the true physical manifold distribution, models can generate diverse samples to comprehensively enrich the feature space \cite{rectifiedflow2023iclr}. This strategy of synergizing generative AI with physical CSI features significantly expands the decision boundaries and has the potential to alleviate the data scarcity bottleneck in HPR systems \cite{multisamplefm2023icml}.

Motivated by these challenges, we propose StructFlow-HPR, a structured pose-conditioned flow matching framework for generative CSI augmentation in 5G HPR systems. The proposed method learns a continuous latent transport process from Gaussian noise to the real CSI manifold under pose guidance, while preserving the physical receiver-frequency structure of CSI measurements. By combining a topology-aware CSI autoencoder with a pose-window conditioned Transformer velocity network, StructFlow-HPR generates pose-aligned CSI samples that can be directly used to augment downstream HPR training under limited-data conditions.

\begin{figure*}[htbp]
\centering
\includegraphics[width=1\linewidth]{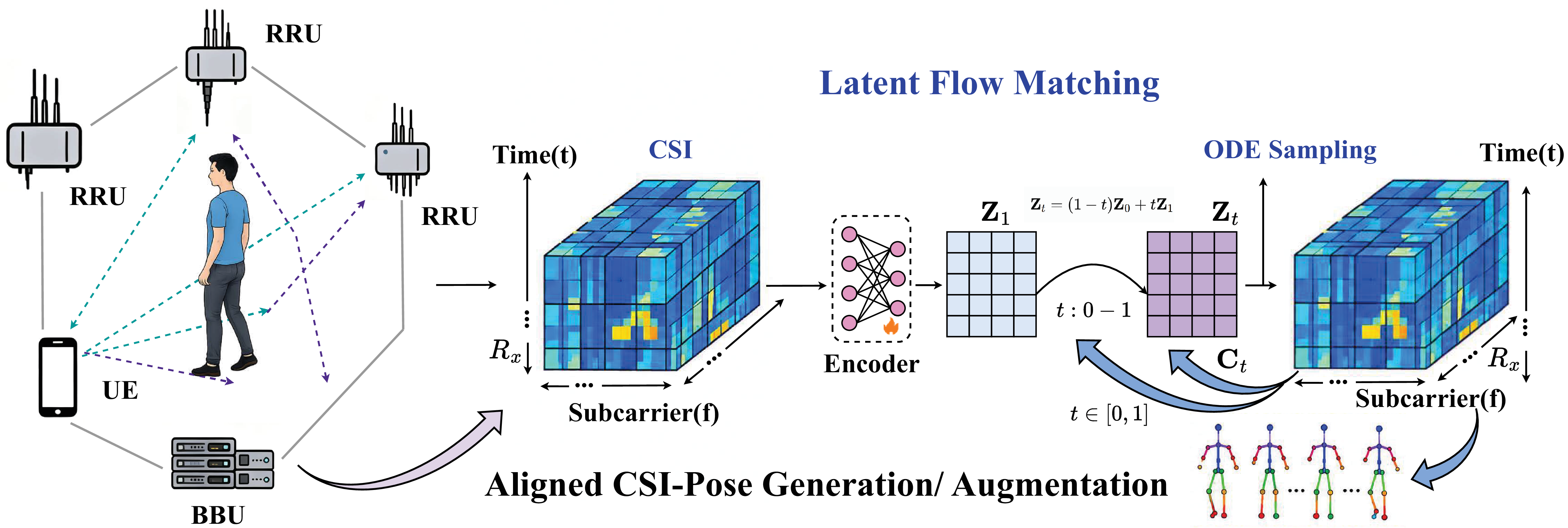}
\caption{Overall framework of StructFlow-HPR for pose-conditioned CSI generation and downstream HPR augmentation.}
\label{fig0}
\end{figure*}

\section{System Model and Problem Formulation}

\subsection{5G Collaborative Sensing and CSI-Pose Representation}

We consider a 5G collaborative sensing system consisting of one transmitting user equipment (UE) and multiple spatially distributed remote radio units (RRUs), which are coordinated by a baseband unit (BBU). The UE transmits uplink sounding reference signals (SRSs), and the RRUs capture the reflected and scattered signal multipath components induced by human motion. As the human body moves in the sensing area, the propagation environment changes accordingly, and such variations are embedded in the measured CSI. 

Let $h(t,k,r)\in\mathbb{C}$ denote the complex CSI at time index $t$, subcarrier index $k$, and receiver index $r$ as
\begin{equation}
    h(t,k,r)=a(t,k,r)e^{j\theta(t,k,r)},
\end{equation}
where $a(t,k,r)$ and $\theta(t,k,r)$ are amplitude and wrapped phase, respectively. To better capture motion-induced variations, we extract multi-domain CSI features from the raw complex measurements.

For the amplitude domain, the temporal amplitude sequence within a sliding window of length $W$ is defined as
\begin{equation}
    \mathcal{A}_{W}(t,k,r)
    =
    \{|h(\tau,k,r)|\}_{\tau=t-W+1}^{t}.
\end{equation}

For $\mathcal{A}_{W}(t,k,r)$, we further compute statistical descriptors such as normalized standard deviation (NSD), median absolute deviation (MAD), and interquartile range (IQR) to characterize local temporal fluctuations:
\begin{equation}
    {\rm NSD}(t,k,r)
    =
    \frac{1}{\mu_A}
    \sqrt{
    \frac{1}{W}
    \sum_{\tau=t-W+1}^{t}
    \left(
    |h(\tau,k,r)|-\mu_A
    \right)^2
    },
\end{equation}
where $\mu_A$ is the mean amplitude in the window.
For phase domain, the wrapped phase $\angle h(t,k,r)$ is unwrapped to obtain a continuous phase $\tilde{\theta}(t,k,r)$. Since the phase difference between receivers is more stable than the absolute phase, we compute the inter-receiver phase difference as
\begin{equation}
    \Delta \tilde{\theta}_{i,j}(t,k)
    =
    \tilde{\theta}(t,k,i)
    -
    \tilde{\theta}(t,k,j).
\end{equation}

Moreover, Doppler-related motion information is derived from temporal phase variation:
\begin{equation}
    f_D(t,k,r)
    =
    \frac{1}{2\pi \Delta t}
    \left(
    \tilde{\theta}(t,k,r)
    -
    \tilde{\theta}(t-1,k,r)
    \right),
\end{equation}
where $\Delta t$ denotes the sampling interval.

After preprocessing, the CSI features of each frame are organized as a structured tensor $\mathbf{X}_t \in \mathbb{R}^{N_c \times N_r \times N_f}$, where $N_c$, $N_r$, and $N_f$ denote the number of subcarriers, receivers, and CSI feature channels, respectively. 

The synchronized human pose label is denoted by $\mathbf{Y}_t \in \mathbb{R}^{J \times K}$, where $J$ is the number of keypoints and $K$ is the coordinate dimension. The aligned CSI-pose dataset is therefore defined as
\begin{equation}
    \mathcal{D}_{\rm raw}
    =
    \{(\mathbf{X}_t,\mathbf{Y}_t)\}_{t=1}^{N_{\rm raw}}.
\end{equation}

\subsection{Problem Formulation for Generative CSI Augmentation}

The downstream HPR task aims to learn a regression model
$f_\phi:\mathcal{X}\rightarrow\mathcal{Y}$ that maps CSI measurements to human poses. Given a limited number of real CSI-pose pairs, the empirical training risk is
\begin{equation}
    \mathcal{R}_{\rm emp}(\phi)
    =
    \frac{1}{N_{\rm raw}}
    \sum_{i=1}^{N_{\rm raw}}
    \mathcal{L}_{\rm HPR}
    \left(
    f_\phi(\mathbf{X}^{(i)}),
    \mathbf{Y}^{(i)}
    \right),
\end{equation}
where $\mathcal{L}_{\rm HPR}(\cdot)$ denotes the pose regression loss. When $N_{\rm raw}$ is small, the learned model tends to overfit the limited training distribution and exhibits poor generalization on unseen real CSI-pose samples.

To mitigate this issue, we formulate data augmentation as conditional CSI generation. Thus, the proposed generator learns the conditional distribution of CSI given pose semantics. Specifically, given pose conditions $\mathbf{Y}$ and latent noise $\boldsymbol{\xi}\sim\mathcal{N}(\mathbf{0},\mathbf{I})$, the generator produces synthetic CSI samples $\hat{\mathbf{X}}$ aligned with the corresponding pose condition. The augmented dataset is defined as
\begin{equation}
    \mathcal{D}_{\rm aug}
    =
    \left\{
    (\hat{\mathbf{X}}^{(j)},\mathbf{Y}^{(j)})
    \right\}_{j=1}^{N_{\rm aug}}.
\end{equation}

The final training set is the union of real and synthetic data,
\begin{equation}
    \mathcal{D}_{\rm mix}
    =
    \mathcal{D}_{\rm raw}
    \cup
    \mathcal{D}_{\rm aug},
\end{equation}

The objective is to learn a pose-conditioned generator that can synthesize realistic CSI-pose pairs and improve the generalization ability of the downstream HPR model under limited-data conditions.

\section{Structured Pose-Conditioned Flow Matching}

\subsection{Structured Latent Representation}

Directly modeling raw CSI tensors is challenging because of their high dimensionality and strong receiver-frequency dependency. To address this issue, StructFlow-HPR first introduces a topology-preserving autoencoder to construct a compact and structured CSI latent space. Let $E_\psi(\cdot)$ and $D_\psi(\cdot)$ denote the encoder and decoder, respectively. For a CSI frame $\mathbf{X}$, the latent code and reconstructed CSI are given by
\begin{equation}
    \mathbf{Z}=E_\psi(\mathbf{X}), \qquad
    \mathbf{X}_{\rm rec}=D_\psi(\mathbf{Z}).
\end{equation}
Moreover, the autoencoder is optimized by minimizing
\begin{equation}
    \mathcal{L}_{\rm AE}(\psi)
    =
    \left\|
    \mathbf{X}
    -
    D_\psi(E_\psi(\mathbf{X}))
    \right\|_1.
\end{equation}

Specifically, the structured latent space preserves the receiver-frequency topology of the wireless measurements, thereby providing a physically meaningful manifold for subsequent generative modeling. 
This design allows the generator to operate on a compact representation while maintaining the topology of the original CSI measurement space.

Human motion is inherently temporal, and a single pose frame is often insufficient to describe the local motion context that causes CSI variations. Therefore, for each target CSI frame $i$, StructFlow-HPR constructs a pose-window condition instead of using only the current pose label. The pose-window condition is defined as
\begin{equation}
    \mathbf{C}_i
    =
    [\mathbf{Y}_{i-r},\ldots,\mathbf{Y}_{i},\ldots,\mathbf{Y}_{i+r}],
    \quad
    r=\frac{w-1}{2},
\end{equation}
where $w$ is the window size. After flattening, the pose-window vector is normalized and embedded by an MLP to obtain a condition representation
\begin{equation}
    \mathbf{h}_Y=E_Y(\mathbf{C}_i).
\end{equation}

Thus, this local strategy captures short-term motion continuity and provides richer semantics than a single pose snapshot. This allows the generator to learn pose-aligned CSI variations that better match the temporal evolution of human movement.

\subsection{Pose-Conditioned Flow Matching and CSI Augmentation}

Given a real CSI latent sample $\mathbf{Z}_1=E_\psi(\mathbf{X})$ and a Gaussian source latent sample $\mathbf{Z}_0\sim\mathcal{N}(\mathbf{0},\mathbf{I})$, StructFlow-HPR defines a linear interpolation path between the source and target latent distributions:
\begin{equation}
    \mathbf{Z}_t
    =
    (1-t)\mathbf{Z}_0
    +
    t\mathbf{Z}_1,
    \qquad
    t\in[0,1].
\end{equation}

The corresponding target velocity is
\begin{equation}
    \mathbf{u}_t
    =
    \frac{d\mathbf{Z}_t}{dt}
    =
    \mathbf{Z}_1-\mathbf{Z}_0.
\end{equation}

A Transformer-based velocity network $v_\theta(\cdot)$ is used to estimate the conditional transport field in the structured latent space. Specifically, CSI latent tokens are processed by self-attention, while the flow time embedding and pose-window embedding modulate the network through adaptive normalization. In this way, the model learns the conditional latent dynamics of CSI evolution under motion guidance rather than predicting diffusion noise. The flow matching objective is formulated as
\begin{equation}
    \mathcal{L}_{\rm FM}(\theta)
    =
    \mathbb{E}_{\mathbf{Z}_0,\mathbf{Z}_1,t,\mathbf{C}_i}
    \left[
    \left\|
    v_\theta(\mathbf{Z}_t,t,\mathbf{C}_i)
    -
    \mathbf{u}_t
    \right\|_2^2
    \right].
\end{equation}

To improve robustness, conditioning dropout can be applied during training so that the network remains stable under both conditional and weakly conditional settings. Compared with diffusion-based generation, this scheme directly learns continuous velocity field in latent space and avoids iterative reverse denoising.
During generation, latent states are initialized from Gaussian noise and evolved according to the learned ordinary differential equation:
\begin{equation}
    \frac{d\mathbf{Z}_t}{dt}
    =
    v_\theta(\mathbf{Z}_t,t,\mathbf{C}_i),
    \qquad
    \mathbf{Z}_0\sim\mathcal{N}(\mathbf{0},\mathbf{I}).
\end{equation}

Using numerical ODE integration, we obtain the generated latent sample $\hat{\mathbf{Z}}$, which is decoded into the CSI domain as
\begin{equation}
    \hat{\mathbf{X}}
    =
    D_\psi(\hat{\mathbf{Z}}).
\end{equation}

Since the generation process is conditioned on the pose window $\mathbf{C}_i$, the generated CSI sample inherits the pose semantics used during sampling and forms a synthetic pair $(\hat{\mathbf{X}},\mathbf{Y})$. Repeating this process yields the augmented dataset
\begin{equation}
    \mathcal{D}_{\rm aug}
    =
    \left\{
    (\hat{\mathbf{X}}^{(j)},\mathbf{Y}^{(j)})
    \right\}_{j=1}^{N_{\rm aug}}.
\end{equation}

Thus, the downstream HPR model is finally trained on the mixed dataset
\begin{equation}
    \mathcal{D}_{\rm mix}
    =
    \mathcal{D}_{\rm raw}
    \cup
    \mathcal{D}_{\rm aug},
\end{equation}
while performance is evaluated exclusively on held-out real CSI-pose samples to verify true generalization rather than memorization of synthetic patterns.

\section{Experiment and Discussions}

\subsection{Experimental Settings}

\subsubsection{5G-enabled Prototype Platform}
To rigorously evaluate the proposed scheme, we construct an indoor multi-node collaborative sensing platform. The 5G experimental infrastructure uses BS equipment from H3C, alongside Sony Xperia 1 IV smartphones serving as the transmitting UE. The core network is driven by the WX3540X, while the BS employs BBU 5200 Series. Spatially, the transmitting smartphone and three receiving RRUs are mounted on tripods and positioned at the four corners of a room, establishing a sensing area of $3.3 \text{ m} \times 2.7 \text{ m}$. All transceivers are fixed at a height of $1.5 \text{ m}$ to optimally capture human kinematic reflections and multipath perturbations within the region.

\begin{table*}[t]
\centering
\caption{Performance comparison between MetaFi, MfDfHPR, and StructFlow-HPR  under various PCK thresholds (Motion States).}
\label{tab:near200_pck_three_methods_delta}
\renewcommand{\arraystretch}{1.2}
\setlength{\tabcolsep}{5pt}
\footnotesize
\begin{tabular}{c|cccc|cccc|cccc|cccc}
\toprule
\multirow{2}{*}{\textbf{State}}
& \multicolumn{4}{c|}{\textbf{PCK5}}
& \multicolumn{4}{c|}{\textbf{PCK10}}
& \multicolumn{4}{c|}{\textbf{PCK20}}
& \multicolumn{4}{c}{\textbf{PCK30}} \\
\cmidrule(lr){2-5}
\cmidrule(lr){6-9}
\cmidrule(lr){10-13}
\cmidrule(lr){14-17}
& \textbf{A1} & \textbf{A2} & \textbf{A3} & $\boldsymbol{\Delta}$
& \textbf{A1} & \textbf{A2} & \textbf{A3} & $\boldsymbol{\Delta}$
& \textbf{A1} & \textbf{A2} & \textbf{A3} & $\boldsymbol{\Delta}$
& \textbf{A1} & \textbf{A2} & \textbf{A3} & $\boldsymbol{\Delta}$ \\
\midrule
Squat
& 67.84 & 76.93 & \textbf{77.79} & +0.86
& 85.67 & 94.42 & \textbf{95.45} & +1.03
& 91.46 & 99.77 & \textbf{99.84} & +0.07
& 91.83 & 100.00 & \textbf{100.00} & +0.00 \\

Move
& 76.61 & 85.78 & \textbf{86.35} & +0.57
& 89.58 & 98.39 & \textbf{99.08} & +0.69
& 91.73 & 99.98 & \textbf{99.95} & -0.03
& 92.16 & 100.00 & \textbf{100.00} & +0.00 \\

Rise Hand1
& 90.39 & 99.22 & \textbf{99.43} & +0.21
& 91.27 & 99.91 & \textbf{99.91} & +0.00
& 92.14 & 100.00 & \textbf{100.00} & +0.00
& 91.76 & 100.00 & \textbf{100.00} & +0.00 \\

Rise Hand2
& 67.18 & 76.31 & \textbf{77.22} & +0.91
& 84.53 & 93.29 & \textbf{94.14} & +0.85
& 90.62 & 99.31 & \textbf{99.31} & +0.00
& 91.49 & 99.84 & \textbf{99.68} & -0.16 \\

Press Leg1
& 43.82 & 52.99 & \textbf{53.63} & +0.64
& 79.06 & 88.03 & \textbf{88.45} & +0.42
& 88.74 & 97.31 & \textbf{97.13} & -0.18
& 88.93 & 97.70 & \textbf{97.82} & +0.11 \\

Press Leg2
& 34.51 & 43.36 & \textbf{47.86} & +4.50
& 65.83 & 74.86 & \textbf{79.99} & +5.13
& 84.27 & 93.45 & \textbf{96.78} & +3.33
& 87.91 & 96.94 & \textbf{98.78} & +1.84 \\
\midrule
Overall
& 63.39 & 72.43 & \textbf{73.71} & +1.28
& 82.66 & 91.48 & \textbf{92.84} & +1.36
& 89.83 & 98.30 & \textbf{98.84} & +0.54
& 90.68 & 99.08 & \textbf{99.38} & +0.30 \\
\bottomrule
\end{tabular}

\vspace{2mm}
\begin{minipage}{0.98\linewidth}
\footnotesize
\emph{Note:} \emph{\textbf{A1 is MetaFi, A2 is MfDfHPR, and A3 is StructFlow-HPR.}} $\Delta$ denotes the performance difference between A3 and A2. Hand1/Hand2 denote Right/Left Hand Raising, and Leg1/Leg2 denote Left/Right Leg Pressing, respectively.
\end{minipage}
\end{table*}

\subsubsection{Data Collection and Augmentation Protocol}

The system synchronously collects 5G uplink SRS-based CSI and visual ground-truth pose sequences for daily human activities, including squatting, hand raising, leg raising, leg pressing, and body translation. The visual pose sequence is temporally aligned with CSI packets according to the nearest timestamp, resulting in frame-level CSI-pose pairs. Each CSI frame is represented by multi-domain wireless features, and each pose frame contains human skeletal keypoints extracted from the synchronized visual stream.

For the prohibitive data acquisition cost of authentic 5G ISAC systems, capturing the raw full-scale data for a single continuous action category inherently consumes approximately 30 minutes of data collection per action category. 
We construct two downstream training settings for comparison, i.e., real-only training using real samples excluding the held-out test segment and real-plus-generated training using all available real training samples and generated samples. All methods are finally evaluated on the same held-out real CSI-pose test set.

\subsubsection{Compared Methods and Evaluation Protocol}
To ensure fairness, we compared state-of-the-art solutions as follows:

\begin{figure*}[t]
\centering
\includegraphics[width=1\linewidth]{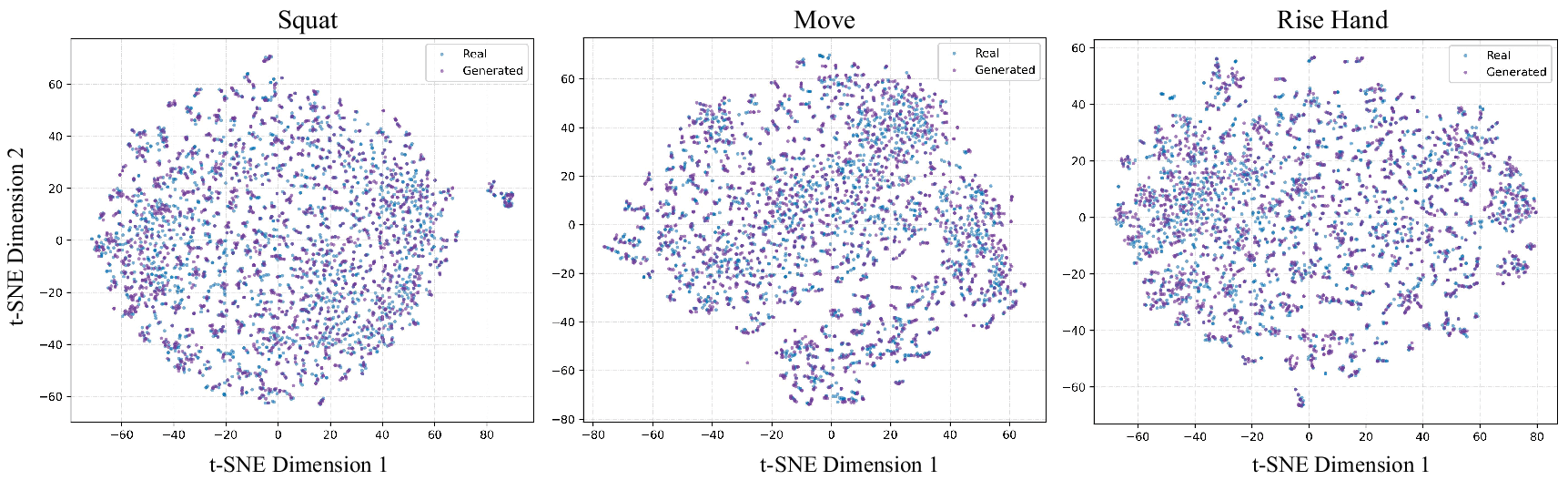}
\caption{t-SNE visualization of structured CSI latent distributions for real and StructFlow-generated samples under representative motion states.}
\label{fig0}
\end{figure*}

\begin{itemize}
    
\item \textbf{MetaFi \cite{10152057}:} A WiFi-enabled IoT human pose estimation scheme originally proposed for metaverse avatar simulation. 
We adapt the MetaFi method to 5G signals by using the same neural network architecture but with our small-sample 5G CSI data as input.

\item \textbf{MfDfHPR \cite{li2025visionwearables5gbased2d}:} As our previously established framework, this model is implemented using PyTorch and trained for 100 epochs to optimize the loss function solely on the original all-sample dataset. 
    
\item \textbf{StructFlow-HPR  (Proposed):} Built upon our previously developed MfDfHPR framework, StructFlow-HPR introduces a structured pose-conditioned flow matching generator for CSI-pose data augmentation. Specifically, the downstream pose regressor strictly adopts the same multi-feature fusion architecture as MfDfHPR, while its training data are augmented with synthetic CSI-pose pairs generated by the proposed flow matching model.
\end{itemize}

To facilitate fair comparisons, the dimensionality of the representations encoded by all methods is standardized. Furthermore, all models utilize the identical neural network backbone originally proposed in MetaFi, ensuring equivalent constraints on computational complexity and memory usage to simulate deployment on resource-limited edge devices.


\begin{figure}[t]
\centering
\includegraphics[width=1\linewidth]{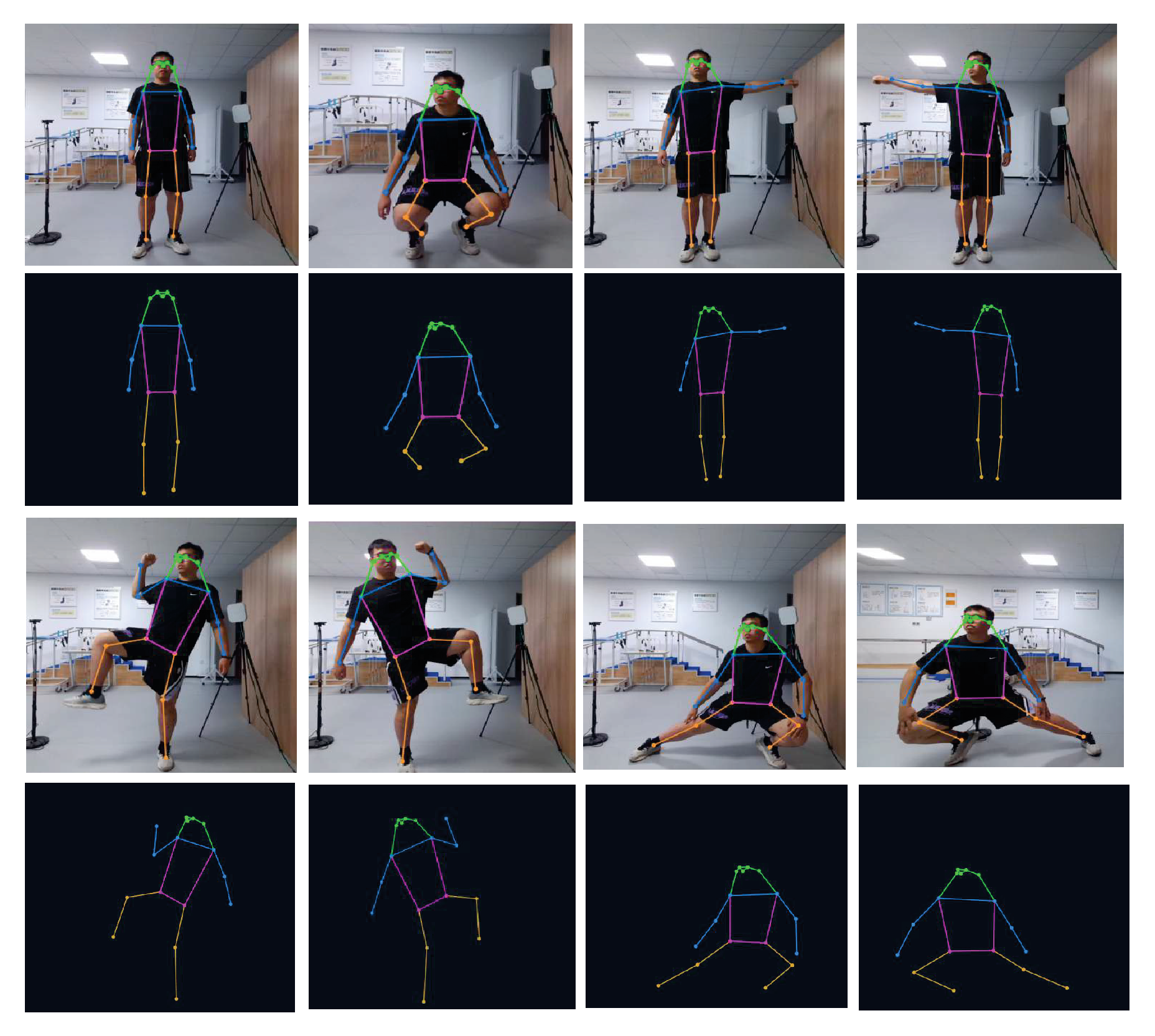}
\caption{The human pose coordinates are generated by the visual model and our proposed StructFlow-HPR  model, respectively.}
\label{fig1}
\end{figure}

\subsection{Experimental Results}

\subsubsection{Generation Quality Evaluation}

Fig.~\ref{fig0} evaluates the generation quality of StructFlow-HPR through latent distribution visualization. Three representative motion states, including Squat, Move, and Rise Hand, are selected for illustration. For each action, real CSI samples and StructFlow-generated CSI samples are first mapped into the structured latent space by the trained autoencoder and then projected into a two-dimensional space using t-SNE. As observed, the generated samples largely overlap with the real samples across different actions, indicating that StructFlow-HPR can learn the underlying pose-conditioned CSI latent distribution rather than producing isolated synthetic points. The close alignment between the two distributions further verifies that the proposed flow matching model preserves the major CSI manifold structure while introducing reasonable sample diversity for downstream HPR augmentation.

\subsubsection{Comparison With Baseline HPR Models}

Table~\ref{tab:near200_pck_three_methods_delta} reports the action-wise HPR performance under different PCK thresholds. A1, A2, and A3 denote MetaFi, MfDfHPR, and StructFlow-HPR, respectively, where StructFlow-HPR  adopts the same pose regression backbone as MfDfHPR but is trained with additional StructFlow-generated CSI-pose samples. PCK$@\alpha$ measures the percentage of correctly predicted keypoints whose error is within $\alpha$\% of the torso length; thus, PCK@5 represents the most stringent setting, while larger thresholds evaluate more relaxed accuracy.

Compared with MetaFi, MfDfHPR achieves consistently higher accuracy on most motion states, confirming the advantage of multi-domain CSI feature modeling. Under the strict PCK@5 criterion, MfDfHPR improves the performance from 67.84\% to 76.93\% on Squat, from 67.18\% to 76.31\% on Rise Hand2, from 43.82\% to 52.99\% on Press Leg1, and from 34.51\% to 43.36\% on Press Leg2. These results indicate that the proposed CSI representation is more effective for fine-grained pose estimation, especially for actions involving large body displacement or limb deformation.

A similar trend can be observed under PCK@10, where MfDfHPR improves the average accuracy from 82.66\% to 91.48\%. For challenging lower-limb actions, the gain is particularly clear: Press Leg1 increases from 79.06\% to 88.03\%, and Press Leg2 increases from 65.83\% to 74.86\%. When the threshold becomes more relaxed, all methods achieve higher PCK values, but MfDfHPR still maintains more stable performance across different motion states. Overall, the average PCK of MfDfHPR reaches 72.43\%, 91.48\%, 98.30\%, and 99.08\% under PCK@5, PCK@10, PCK@20, and PCK@30, respectively, demonstrating its robustness for 5G-based HPR.

\subsubsection{Real-World Validation of StructFlow Augmentation}

With StructFlow-based augmentation, StructFlow-HPR  further improves the overall performance over MfDfHPR. The average PCK increases from 72.43\% to 73.71\% at PCK@5, from 91.48\% to 92.84\% at PCK@10, from 98.30\% to 98.84\% at PCK@20, and from 99.08\% to 99.38\% at PCK@30. The improvement is particularly evident on challenging lower-limb actions. For Press Leg2, StructFlow-HPR  improves PCK@5 by 4.50\%, PCK@10 by 5.13\%, and PCK@20 by 3.33\%. These results demonstrate that StructFlow-HPR can generate useful pose-aligned CSI variations that complement the real training data and enhance the generalization ability of downstream HPR models under limited-data conditions.

\section{Conclusions and Future Work}

In this paper, we propose StructFlow-HPR, a structured pose-conditioned flow matching framework for generative 5G CSI augmentation. By preserving the receiver-frequency topology of CSI signals and learning pose-guided latent transport, the proposed method generates CSI samples aligned with human pose labels and improves downstream HPR performance under limited-data conditions. Future work will focus on more adaptive action-specific augmentation strategies and extension to cross-environment and multi-person 5G sensing scenarios.

\bibliographystyle{IEEEtran} 
\bibliography{bib}

\begin{thebibliography}{10}
\providecommand{\url}[1]{#1}
\csname url@samestyle\endcsname
\providecommand{\newblock}{\relax}
\providecommand{\bibinfo}[2]{#2}
\providecommand{\BIBentrySTDinterwordspacing}{\spaceskip=0pt\relax}
\providecommand{\BIBentryALTinterwordstretchfactor}{4}
\providecommand{\BIBentryALTinterwordspacing}{\spaceskip=\fontdimen2\font plus
\BIBentryALTinterwordstretchfactor\fontdimen3\font minus \fontdimen4\font\relax}
\providecommand{\BIBforeignlanguage}[2]{{%
\expandafter\ifx\csname l@#1\endcsname\relax
\typeout{** WARNING: IEEEtran.bst: No hyphenation pattern has been}%
\typeout{** loaded for the language `#1'. Using the pattern for}%
\typeout{** the default language instead.}%
\else
\language=\csname l@#1\endcsname
\fi
#2}}
\providecommand{\BIBdecl}{\relax}
\BIBdecl

\bibitem{10944626}
H.~Zhang, Z.~Zhang, X.~Liu, W.~Li, H.~Li, and C.~Sun, ``Integrated sensing and communication for {6G} holographic {D}igital {T}wins,'' \emph{IEEE Wireless Communications}, vol.~32, no.~2, pp. 104--112, Mar. 2025.

\bibitem{vitposepp2024tpami}
Y.~Xu, J.~Zhang, Q.~Zhang, and D.~Tao, ``Vitpose++: Vision transformer for generic body pose estimation,'' \emph{IEEE Trans. Pattern Anal. Mach. Intell.}, vol.~46, no.~2, pp. 1212--1230, Feb. 2024.

\bibitem{dwpose2023iccvw}
Z.~Yang, A.~Zeng, C.~Yuan, and Y.~Li, ``Effective whole-body pose estimation with two-stages distillation,'' in \emph{IEEE Int. Conf. Comput. Vis. Workshops (ICCVW), Paris, France}, Oct. 2023, pp. 4210--4220.

\bibitem{humanart2023cvpr}
X.~Ju and et~al., ``Human-art: A versatile human-centric dataset bridging natural and artificial scenes,'' in \emph{IEEE Conf. Comput. Vis. Pattern Recognit. (CVPR), Vancouver, Canada}, June 2023, pp. 618--629.

\bibitem{personinwifi3d2024cvpr}
K.~Yan and et~al., ``Person-in-{W}i{F}i {3D}: End-to-end multi-person {3D} pose estimation with {W}i-{F}i,'' in \emph{IEEE Conf. Comput. Vis. Pattern Recognit. (CVPR), Seattle, WA, USA}, June 2024, pp. 969--978.

\bibitem{11145172}
A.~Ghosh, T.~Wild, J.~Du, J.~Tan, A.~Grudnitsky, D.~Chizhik, S.~Mandelli, Y.~Xing, F.~Schaich, and H.~Viswanathan, ``A unified future: Integrated sensing and communication (isac) in 6g,'' \emph{IEEE J. Sel. Top. Electromagn. Antennas Antennas}, vol.~1, no.~1, pp. 365--374, Aug. 2025.

\bibitem{capshar2024jiot}
R.~Djogo and et~al., ``Fresnel zone-based voting with capsule networks for human activity recognition from channel state information,'' \emph{IEEE Internet Things J.}, vol.~11, no.~13, pp. 23\,309--23\,321, Apr. 2024.

\bibitem{10152057}
J.~Yang and et~al., ``Meta{F}i: Device-free pose estimation via commodity {W}i{F}i for {M}etaverse avatar simulation,'' in \emph{IEEE World Forum Internet Things (WF-IoT), Hybrid, Yokohama, Japan}, Oct. 2022, pp. 1--6.

\bibitem{isac_standardization2024mcomstd}
A.~Kaushik, R.~Singh, S.~Dayarathna, R.~Senanayake, M.~Di~Renzo, M.~Dajer, H.~Ji, Y.~Kim, V.~Sciancalepore, A.~Zappone, and W.~Shin, ``Toward integrated sensing and communications for {6G}: Key enabling technologies, standardization, and challenges,'' \emph{IEEE Commun. Stand. Mag.}, vol.~8, no.~2, pp. 52--59, May 2024.

\bibitem{diffusionts2024iclr}
X.~Yuan and Y.~Qiao, ``Diffusion-{TS}: Interpretable diffusion for general time series generation,'' in \emph{Int. Conf. Learn. Represent. (ICLR), Vienna Austria}, vol. 2024, May 2024, pp. 41\,582--41\,610.

\bibitem{dit2023iccv}
W.~Peebles and S.~Xie, ``Scalable diffusion models with transformers,'' in \emph{IEEE Int. Conf. Comput. Vis. (ICCV)}, Oct. 2023, pp. 4172--4182.

\bibitem{flowmatching2023iclr}
Y.~Lipman, R.~T.~Q. Chen, H.~Ben-Hamu, M.~Nickel, and M.~Le, ``Flow matching for generative modeling,'' in \emph{Int. Conf. Learn. Represent. (ICLR), Kigali Rwanda}, May 2023.

\bibitem{rectifiedflow2023iclr}
X.~Liu, C.~Gong, and Q.~Liu, ``Flow straight and fast: Learning to generate and transfer data with rectified flow,'' in \emph{Int. Conf. Learn. Represent. (ICLR), Kigali Rwanda}, May 2023.

\bibitem{multisamplefm2023icml}
A.-A. Pooladian, H.~Ben-Hamu, C.~Domingo-Enrich, B.~Amos, Y.~Lipman, and R.~T.~Q. Chen, ``Multisample flow matching: Straightening flows with minibatch couplings,'' in \emph{Proc. Int. Conf. Mach. Learn. (ICML), Honolulu, USA}, July 2023, pp. 28\,100--28\,127.

\bibitem{li2025visionwearables5gbased2d}
H.~Li, D.~Li, A.~Zhang, W.~Zhang, C.~Sun, and H.~Zhang, ``No vision, no wearables: {5G}-based {2D} human pose recognition with integrated sensing and communications,'' \emph{arXiv.cs.CV.2512.24923}, Dec. 2025.

\end{thebibliography}

\end{document}